\documentclass[nohyper,12pt,letterpaper]{JHEP3}

\usepackage{amsmath,epsfig}
\usepackage[latin1]{inputenc}
\usepackage{amssymb,amsfonts}

\title{Higher Derivative Corrections, \\
Dimensional Reduction and Ehlers Duality}

\preprint{{\tt arXiv:0706.1769}\\LPTHE-07\\LPTENS-07-24}

\author{Yann Michel$^{\rm a}$ and Boris Pioline$^{\rm a, b}$\\

$^a$
Laboratoire de Physique Th\'eorique et Hautes Energies\footnote{Unit\'e mixte
de recherche du CNRS UMR 7589}, \\
Universit\'e Pierre et Marie Curie - Paris 6,
4 place Jussieu, F-75252 Paris cedex 05 \\

$^b$ Laboratoire de Physique Th\'eorique de l'Ecole Normale
Sup\'erieure\footnote{Unit\'e mixte
de recherche du CNRS UMR 8549}\\
24 rue Lhomond, F-75231 Paris cedex 05\\

\medskip

{\tt E-mail: ymichel@clipper.ens.fr, pioline@lpthe.jussieu.fr} }

\abstract{Motivated by applications to black hole physics and duality,
we study the effect of higher derivative corrections on the dimensional
reduction of four-dimensional Einstein, Einstein-Liouville and
Einstein-Maxwell gravity
to one direction, as appropriate for stationary, spherically symmetric
solutions. We construct a field redefinition scheme such that the
one-dimensional Lagrangian is corrected only by powers of first derivatives
of the fields, eliminating spurious modes and providing a suitable
starting point for quantization. We show that the Ehlers symmetry,
broken by the leading $R^2$ corrections in Einstein-Liouville
gravity, can be restored by including
contributions of Taub-NUT instantons. Finally, we give a preliminary
discussion of the duality between higher-derivative F-term corrections on
the vector and hypermultiplet branches in $N=2$ supergravity in four
dimensions.}

\newcommand{\pa}{\partial}
\newcommand{\nn}{\nonumber}

\newcommand{\IR}{\mathbb{R}}

\newcommand{\IZ}{\mathbb{Z}}

\newcommand{\tzeta}{{\tilde\zeta}}

\newcommand{\cM}{\mathcal{M}}

\def\m{\mu}

\def\e{\epsilon}

\def\bea{\begin{eqnarray}}
\def\eea{\end{eqnarray}}
\def\be{\begin{equation}}
\def\ee{\end{equation}}
\def\ba{\begin{align}}
\def\ea{\end{align}}
\def\bse{\begin{subequations}}
\def\ese{\end{subequations}}
\def\1F1{{}_1\!F_1}
\def\2F0{{}_2\!F_0}

\begin{document}


\section{Introduction and Summary}

In the study of stationary, spherically symmetric solutions to
Einstein's gravity, possibly coupled to Maxwell and massless scalar fields,
a useful trick is to first dimensionally reduce the action
along the time-like Killing vector, and then only to enforce spherical
symmetry of the three-dimensional spatial slices \cite{Breitenlohner:1987dg}.
The advantage of this two-step procedure is that
three-dimensional gauge fields arising in the reduction can be dualized
after the first step into
pseudo-scalars, leading to a non-linear sigma model with target
space $\cM_3^*$, coupled to three-dimensional gravity.
The second step leads to a one-dimensional
dynamical system describing the geodesic motion of a fiducial particle
on the cone $\IR^+\ltimes \cM_3^*$. The factor $\IR^+$ describes
the radius of the two-sphere, while $\cM_3^*$ is an analytic
continuation of the moduli space $\cM_3$ arising  in the
usual Kaluza-Klein reduction along a space-like direction, leading
to a metric with indefinite signature. In many cases, $ \cM_3^*$ has
a larger group $G_3$ of non-compact symmetries beyond those already manifest
in four dimensions ($G_4$), which may allow to integrate the geodesic
motion explicitly.

The simplest examples where this procedure has been useful are
Schwarzschild-NUT black holes in pure Einstein gravity, which
correspond to geodesics on the Poincar\'e upper half plane
$Sl(2,\IR)/SO(2)$ \cite{Geroch:1970nt}. The $Sl(2,\IR)$ action on
this space, often known as Ehlers' symmetry, relates static
solutions with solutions with non-zero NUT charge
\cite{Geroch:1970nt,Ehlers}. Similarly, Reissner-Nordstr\"om-NUT
black holes in Einstein-Maxwell supergravity correspond to
geodesics on $SU(2,1)/Sl(2)\times U(1)$~\cite{Kinnersley:1977pg,
Mazur:1983dc} (see also the review \cite{Pioline:2006ni} on this
and other issues to be discussed below). The reduction of
four-dimensional Einstein-Maxwell theories coupled to scalars
valued in a symmetric space $G_4/K_4$ was worked out in
\cite{Breitenlohner:1987dg,Breitenlohner:1998cv}, leading to a
non-linear sigma model on $G_3/K_3$, whose relation to $G_4/K_4$
can be most easily expressed using the language of Jordan algebras
\cite{Gunaydin:2005df}. These theories are bosonic truncations of
a special class of ${\cal N}=2$ supergravity theories with
symmetric moduli spaces~\cite{Gunaydin:1983bi}.
More generally, the reduction of ${\cal
N}=2$ supergravity coupled to $n_V$ Abelian vector multiplets
leads to a non-linear sigma model on a para-quaternionic-K\"ahler
manifold $\cM_3$ of dimension $4n_V+4$, known as the $c^*$-map of
the four-dimensional moduli space $\cM_4$
\cite{Cecotti:1988qn,Ferrara:1989ik, Gunaydin:2005mx}. For a
particular class of geodesics, corresponding to BPS black holes,
the motion can be fully integrated (both classically and quantum
mechanically) \cite{Neitzke:2007ke}, and reproduces the usual
attractor flow equations controlling the radial evolution of the
scalars on $\cM_4$ \cite{Gunaydin:2005mx,Pioline:2006ni}.

In all of these examples, the starting point was Einstein-Hilbert gravity in
four dimensions coupled to abelian gauge fields and scalar fields,
with canonical two-derivative kinetic terms for all fields.
In general however, there are higher-derivative corrections to the
four-dimensional effective action coming from integrating out massive modes
in the full quantum theory, suppressed by inverse
powers of the Planck mass $m_P$. A prime example are the $R^2$ corrections,
which play an essential role in accounting for the microscopic entropy
of ``small black holes'', whose horizon is singular in the two-derivative
approximation: such corrections become dominant near the singularity, and
lead to a smooth near-horizon geometry in agreement with thermodynamical
expectations~\cite{Dabholkar:2004yr,Dabholkar:2004dq,
Sen:2004dp,Hubeny:2004ji,Dabholkar:2005by,Dabholkar:2005dt,Pioline:2006ni}.
Higher derivative corrections are also important for regular black holes
when trying to account for finite size corrections to the thermodynamical
limit.

The main goal of this work is to analyze the effect of such higher derivative
gravitational corrections at the level of the dimensional reduction from
four space-time dimensions to one radial dimension.
Our approach is also suitable for analyzing higher-derivative
corrections to the usual Kaluza-Klein reduction from four to three
space-time dimensions, as we discuss further below. On general grounds,
one expects higher-derivative corrections to the geodesic motion
on $\IR^+ \times \cM_3$, preserving only the symmetries which originate
from gauge and diffeomorphism invariance in four dimensions.
As usual when working beyond two-derivative order, the
exact form of the higher-derivative corrections is largely ambiguous
due to the freedom of performing field redefinitions. A preferred
frame is one in which only powers of {\it first}
derivatives of the scalars appear, as it removes spurious modes
found in other schemes \cite{Sen:2004dp,Hubeny:2004ji}, and
is amenable to canonical quantization by standard means. Moreover, 
it is also the natural frame in which to assess the existence of hidden 
symmetries, as will become clear shortly.

As the first part of our investigation, we consider
pure Einstein gravity in four dimensions, with an arbitrary combination
of the three curvature-squared invariants, and study its
dimensional reduction on stationary spherically symmetric geometries.
We show that all higher-derivative corrections in the one-dimensional
Lagrangian can be removed by suitable field redefinitions.
This could be anticipated from the fact
that $R^2$ corrections in four dimensions can always be related to
the Gauss-Bonnet density by an appropriate field redefinition of
the four-dimensional graviton~\cite{Zwiebach:1985uq,Colonnello:2007qy}.
This is no longer true in the case of Einstein gravity coupled
to a scalar field $\phi$, which we study next: allowing the coefficients
of the $R^2$ terms to depend on $\phi$, we find that there exists
a frame where only powers of first derivatives of the fields
appear (with always at least one power of the first derivative of $\phi$, 
in agreement with the triviality of $R^2$ corrections with constant 
coefficients).
Finally, we perform a similar
analysis for Einstein gravity coupled to a single Maxwell field with general
four-derivative couplings (but restricting for simplicity to the
static case, rather than stationary). We find that higher-derivatives
cannot in general be eliminated by field redefinitions, but
that a first order scheme can still be found. We note that extremal solutions
exist only when the coefficient of $(F_{\mu\nu}^2)^2$ vanishes,
which may originate from the possibility of supersymmetrizing
the higher-derivative corrections.

As indicated above, our analysis also addresses the effect of higher-derivative
corrections on the usual Kaluza-Klein reduction on a space-like direction,
after analytic continuation of the scalar fields in the Maxwell directions.
Of course, since we perform the reduction in the spherically symmetric sector
only, the higher-derivative corrections in three-dimensions are only
determined up to some tensor structure ambiguities.

In this context, an interesting question is whether higher-derivative
interactions can be consistent with the extended non-compact
symmetries ($G_3$) that were present at tree-level, or at least with
a discrete subgroup thereof. This question
was raised long ago in the context
of T-duality \cite{Meissner:1996sa,Kaloper:1997ux}, where it was found
that $\alpha'$ corrections to the T-duality rules could be reabsorbed
by field redefinitions, leaving an action invariant under $R\to 1/R$.
In the context of U-duality, the same question arises as to whether
higher-derivative interactions, beyond the already well understood
gravitational sector, preserve the duality symmetry
$G_3(\IZ)=E_{8(8)}(\mathbb{Z})$ of M-theory compactified
on a eight-torus~\cite{Lambert:2006he,Lambert:2006ny}.
Since an $Sl(2,\mathbb{R})$ subgroup of $G_3$ originates in the
Ehlers symmetry of the reduction of four-dimensional Einstein gravity to three
dimensions, we can investigate a toy version of this problem, and
ask whether a discrete subgroup of the Ehlers symmetry can be preserved
by $R^2$ corrections. For pure Einstein gravity, the answer
to this question is trivial since such corrections can always be removed
by field redefinitions. We therefore address this problem in the context
of Einstein-Liouville theory, where such corrections are non-trivial,
and break the Ehlers symmetry at order $\alpha'$.
Having found a frame where only powers of the first order
derivatives of the scalar fields appear, it is straightforward
to restore the invariance under a discrete subgroup
$Sl(2,\IZ)$ of Ehlers symmetry: after
upon expressing the Lagrangian in
powers of the right-invariant form $p$ in $Sl(2)/U(1)$ and its
complex conjugate $\bar p$, with weight one under $U(1)$,
it suffices to replace the coefficient of any term
proportional to $p^m {\bar p}^n$ by a generalized
Eisenstein series $f_{s,k=m-n}$ with $U(1)$ weight $k=m-n$, of the
type considered in \cite{Green:1997me,Kehagias:1997jg}. For
$s=1$, the case relevant for $R^2$ corrections, $f_{s,k}$ can in
fact be expressed in terms of ordinary almost holomorphic modular
forms.  The difference between the $Sl(2,\IZ)$-invariant
and the original dimensionally reduced Lagrangian can be attributed to
Taub-NUT gravitational instantons, as well as
loops of gravitons running along the compact
circle. Thus, we give a precise realization of the general
expectations expressed in~\cite{Lambert:2006he,Lambert:2006ny}.

Finally, a third motivation for our work is to further our
understanding of the duality between hypermultiplets and vector
multiplets in three dimensions beyond the two-derivative level. In
type II string theory compactified on a Calabi-Yau three-fold $Y$
times a circle, T-duality along the circle exchanges the
hypermultiplet and vector multiplet moduli spaces of the type IIA and
type IIB theories, respectively. Since the hypermultiplet sector
is independent of the size of the circle, and since the metric for
the three-dimensional vector multiplets is given by the $c$-map of
the four-dimensional vector multiplet metric, this implies that,
at tree-level, the four-dimensional hypermultiplet space is given
by the $c$-map of the four-dimensional vector multiplet space of
the dual theory~\cite{Cecotti:1988qn,Ferrara:1989ik}. The hypermultiplet
metric is further corrected by
D-instantons, dual to black holes winding around
the circle on the vector-multiplet side. It is natural to guess
that the same argument should relate the $R^2 F^{2h-2}$
``F-term'' higher derivative corrections  on the
vector multiplet side in four dimensions to the
$(\nabla^2 S)^2 (\nabla Z)^{2h-2}$ ``$\tilde F$-term''
higher-derivative corrections
on the hypermultiplet side. Here $(S,Z)$
denotes the two chiral fields of the universal hypermultiplet
\cite{Antoniadis:1993ze}. In Section \ref{cmaphd}, we give a
preliminary analysis of this problem in the simplest case with
$h=1$ and $(n_V,n_H)=(0,1)$, and conclude that the identication
between $F_1$ and $\tilde F_1$ is more subtle than commonly
thought.

The organization of this paper is as follows. In Section 2, we
discuss the reduction of four-dimensional Einstein,
Einstein-Liouville and Einstein-Maxwell
gravity with four-derivative corrections to one radial dimension,
and find suitable field redefinitions such that the resulting Lagrangian
involves only powers of first derivatives of the fields. In Section 3,
we discuss the restoration of Ehlers symmetry via
instanton corrections. In Section 4, we give a preliminary discussion
of the relation between the higher-derivative F-term couplings
$F_1$ and $\tilde F_1$  on the vector and hypermultiplet branch
in three dimensions.

As this work was finalized, we received \cite{Bao:2007er}, which
has some overlap with the results in Section 3.
{\it Note added:} After the first version of this
paper appeared on the archive, the authors of \cite{Colonnello:2007qy}
pointed out that $R^2$ corrections in pure Einstein gravity can be completely
removed by field redefinitions, not just in the sector with flat sections
as we erroneously claimed. In this revised version,
we extend our discussion to the case of Einstein-Liouville gravity,
where $R^2$ corrections are non-trivial and do break the Ehlers symmetry.

\section{Spherical Reduction and Higher-Derivative Terms}

\subsection{Pure Einstein Gravity} \label{einstein}

In this Section, we study pure gravity in four dimensions,
with four-derivative corrections to the Einstein-Hilbert action:
\begin{equation}
\label{S4}
S = \int d^4x\,\sqrt{-g_4}\,\left[R^{(4)} +
\alpha\,([R^{(4)}_{\mu\nu}]^2-[R^{(4)}]^2) + \beta \,[R^{(4)}]^2
+\gamma\, R_{GB}^2 + o(\alpha')
\right]
\end{equation}
and $o(\alpha')$ denotes further higher derivative corrections
derivatives, which we assume to be negligible compared to the four-derivative
interactions displayed in \eqref{S4}. Throughout this paper, we work
perturbatively in $(\alpha,\beta,\gamma)\sim \alpha'$. Since
the Gauss-Bonnet density $R_{GB}^2=[R^{(4)}_{\mu\nu\rho\sigma}]^2
-4[R^{(4)}_{\mu\nu}]^2 + [R^{(4)}]^2$ is a total derivative in four
dimensions, we set $\gamma=0$ in this section.
The action \eqref{S4} can then be specialized
to the Riemann tensor squared $[R^{(4)}_{\mu\nu\rho\sigma}]^2$ or
the Weyl tensor squared $[W^{(4)}_{\mu\nu\rho\sigma}]^2$ forms by setting
$\alpha=3\beta/4$ or $\alpha=\beta/6$, respectively.

In the presence of a time-like Killing vector, the four-dimensional
metric may be written as
\be
\label{ds4}
ds_4^2 = - e^{2U} ( dt + \omega )^2 + e^{-2U} ds^2_3
\ee
where the scalar $U$, Kaluza-Klein one-form $\omega$ and spatial
three-dimensional metric $ds^2_3$ are independent of the time
coordinate $t$. The action \eqref{S4} may then be reduced along the
ansatz \eqref{ds4}, leading to
\be
S_3 = \int d^3x \sqrt{g_3} \left[ R^{(3)} -2
(\pa_i U)^2 + \frac{1}{4} e^{2U} F_{ij}^2 + {\cal O}(\alpha') \right]
\ee
where $F_{ij}=\pa_i \omega_j - \pa_j \omega_i$. The choice of powers
of $e^{U}$ in \eqref{ds4} ensures that the three-dimensional
action is obtained in the Einstein frame. The terms ${\cal O}(\alpha')$
coming from the reduction of the four-derivative terms in \eqref{S4}
are somewhat cumbersome to obtain. For simplicity, and motivated
by application to black holes, we further restrict to
spherically symmetric solutions,
\begin{equation}
\label{ds4sph}
ds_4^{2} = -e^{2U}(dt+k\cos{\theta}\,d\phi)^2 + e^{-2U}\left[
N^2(\rho)\, d\rho^2 +
r^2(\rho)\, (d\theta^2+\sin^{2}{\theta}\,d\phi^{2})\right]
\end{equation}
The integer $k$, often known as the NUT charge, describes
the first Chern class of the Kaluza-Klein gauge field over the two-sphere
at infinity; it can be dualized to a three-dimensional scalar $\sigma$
by adding a Lagrange multiplier $k \sigma'$ to the 
action\footnote{In the following the 
primes denote $\rho$-derivatives.}, which
ensures that $k$ is a constant of motion. It is worth emphasizing that the
ansatz \eqref{ds4sph} follows entirely from the assumed isometries,
in particular the Kaluza-Klein connection $k \cos\theta\, d\phi$ is
unaffected by higher-derivative corrections to the Einstein-Hilbert
action, and $\sigma'$ will continue to be related to $k$ by
Legendre transform. 
The lapse variable $N(\rho)$ can be viewed as an einbein
along the radial direction $\rho$, and ensures that the resulting
one-dimensional action is invariant under diffeomorphisms of $\rho$.
For convenience we shall often set $N(\rho)=1$, the dependence on
$N$ can be reinstated whenever needed by demanding reparametrization 
invariance.

It is now straightforward, if tedious, to compute the curvature invariants of
the metric \eqref{ds4sph}, and integrating over the sphere coordinates
$\theta,\phi$ to obtain the one-dimensional Lagrangian. At two-derivative
order, we find the familiar tree-level result
\bea
\mathcal{L}_0 &=&  2N \left[ \left(\frac{r'}{N}\right)^2
- r^2\, \left(\frac{U'}{N}\right)^2 +\frac{e^{4U}}{4r^2}k^2 + 1 \right]
+  k \sigma' \\
&=&  \frac{2}{N} \left[ r'^2- r^2 \left( U'^2 + \frac14
e^{-4U} \sigma'^2 \right)
+ N^2 \right]
\eea
where in the second line we have performed the Legendre transform
over the NUT charge $k$.
The term in bracket is recognized as the metric on the cone
$\IR^+ \times Sl(2,\IR)/U(1)$ where
\be
\label{deftau}
\tau=\sigma+ i e^{2U}
\ee
is the standard
coordinate on the upper half-plane. Stationary, spherically symmetric
solutions of Einstein gravity in four dimensions are therefore described
by the geodesic motion of a fiducial particle with unit mass
on $\IR^+ \times Sl(2,\IR)/U(1)$. The $Sl(2,\IR)$ symmetry
acting on $\tau$ by fractional linear transformations
\be
\label{sl2eh}
\tau \to \frac{a\tau+b}{c\tau+d} \ ,\quad
\ee
is Ehlers's symmetry mentioned in the introduction.

Including the four-derivative interactions, and setting $N=1$,
we arrive at $\mathcal{L}=\mathcal{L}_0+\mathcal{L}_1$ where
\begin{eqnarray}
\mathcal{L}_1 & =   & (-2\alpha+4\beta)\frac{e^{2U}}{r^2}
            +(8\alpha-8\beta)e^{2U}U'^2
            +(-4\alpha+8\beta)e^{2U}U''  \\
            &   & +(4\alpha-8\beta)\frac{e^{2U}r'^2}{r^2}
            +(-8\alpha+16\beta)\frac{e^{2U}r'U'}{r}
            +(12\alpha-16\beta)\frac{e^{2U}r''}{r} \nonumber \\
            &   & +(-10\alpha+16\beta)e^{2U}r''^2
            +(-2\alpha+4\beta)\frac{e^{2U}r'^4}{r^2}
            +(16\alpha-32\beta)e^{2U}r'r''U' \nonumber \\
            &   & +(-8\alpha+24\beta)e^{2U}r'^2U'^2
            +(8\alpha-16\beta)e^{2U}rr'U'^3
            +(4\alpha-8\beta)e^{2U}r^2U'^2U''  \nn \\
            &   & +(-12\alpha+16\beta)\frac{e^{2U}r'^2r''}{r}
            +(8\alpha-16\beta)e^{2U}rr''U'' \nonumber  \\
            &   & +(4\alpha-8\beta)e^{2U}r'^2U''
            +16\beta e^{2U}rr'U'U''  \nonumber  \\
     &   & +(-8\alpha+16\beta)e^{2U}rr''U'^2
            +(8\alpha-16\beta)\frac{e^{2U}r'^3U'}{r}
            +2\beta e^{2U}r^2U'^4+4\beta e^{2U}r^2U''^2\nonumber \\
            &   & +(2\alpha-4\beta)k^2\frac{e^{6U}r''}{r^3}
            +(2\alpha+4\beta)k^2\frac{e^{6U}r'U'}{r^3}
            +(\alpha+2\beta)k^2\frac{e^{6U}U''}{r^2} \nonumber \\
            &   & +(2\alpha-2\beta)k^2\frac{e^{6U}U'^2}{r^2}
            +2\beta k^2\frac{e^{6U}}{r^4} -2\beta
            k^2\frac{e^{6U}r'^2}{r^4}
       +\left(\frac{1}{2}\alpha+\frac{1}{4}\beta\right)k^4\frac{e^{10U}}{r^6}
       \nonumber
\end{eqnarray}
It is worth noting that ${\cal L}_0$ and ${\cal L}_1$
are homogeneous, of degree 0 and $-2$ respectively, under the global symmetry
\begin{equation}
        \rho \to e^{2l} \, \rho\ ,\quad
        U(\rho) \to U(\rho) + l\ ,\quad
        r(\rho) \to e^{2l}\,r(\rho)\ ,\quad
        k \to k
\label{dilatation}
\end{equation}
This reflects the homogeneity of the Einstein-Hilbert and $R^2$
terms, respectively, under global rescaling $g_{\mu\nu}\to e^{2l} g_{\mu\nu}$.

The Lagrangian $\mathcal{L}$ should be supplemented by the
Hamiltonian constraint, or Wheeler-De Witt equation, coming from
the  equation of motion of $N$.  The latter can be
reinstated by replacing all derivatives with respect to $\rho$ by
covariant derivatives $\nabla_\rho$ with respect to the world-line
metric $\gamma_{\rho\rho}= N^2$, contracted with appropriate
powers of the inverse metric. This task is greatly simplified if
one first performs field redefinitions and integration by parts
such that the resulting action only involve powers of first order
derivatives of $U(\rho)$ and $r(\rho)$. The most general
redefinition\footnote{Additional first-order terms proportional to
$e^{2U}r'/r$ and $e^{2U}U'/r$ would spoil
one-dimensional diffeomorphism invariance and are therefore
not considered. Moreover, we do not allow for field redefinitions
of $k$, since $k$ corresponds to a conserved charge.} is
\bea \delta r &=&
\left(-\frac{5}{2}\alpha+4\beta\right)e^{2U}r''+(-\alpha+2\beta)e^{2U}r'U'+(2\alpha-4\beta)e^{2U}rU'' \nonumber \\
&& +\left(\frac{1}{4}\alpha-\beta\right)e^{2U}rU'^2 +
x_1\frac{e^{2U}}{r}+x_2\frac{e^{2U}r'^2}{r}
+x_3\frac{e^{6U}}{r^3}k^2 \label{FRgen} \\
\delta U &=&
\left(\frac{3}{2}\alpha-3\beta\right)\frac{e^{2U}r'^2}{r^2}
+\frac{1}{2}\alpha\,\frac{e^{2U}r'U'}{r}-\beta
e^{2U}U'' \nonumber \\
&& + y_1\frac{e^{2U}}{r^2}+y_2\e^{2U}U'^2
+y_3\frac{e^{6U}}{r^4}k^2 \nonumber \eea where
$x_1,x_2,x_3,y_1,y_2,y_3$ are six arbitrary parameters, which can
be chosen at will to simplify the form of the final Lagrangian.
After dropping a total derivative, the Lagrangian $\mathcal{L}$
becomes
\begin{eqnarray}
\mathcal{L} & = & 2+2r'^2-2r^2U'^2+\frac{e^{4U}}{2r^2}k^2 + k \sigma'
\label{l1gen} \\
            &   & +(-2\alpha+4\beta)\frac{e^{2U}}{r^2}
            +(16\alpha-24\beta-4x_1)\frac{e^{2U}r'^2}{r^2} \nonumber \\
            &   & +(-32\alpha+48\beta+8x_1+8y_1)\frac{e^{2U}r'U'}{r}
            +(16\alpha-24\beta-4x_1-8y_1)e^{2U}U'^2 \nonumber \\
            &   & +\left(-6\alpha+28\beta-\frac{4}{3}x_2\right)\frac{e^{2U}r'^4}{r^2}
            +
            \left(28\alpha-\frac{152}{3}\beta+\frac{8}{3}x_2\right)\frac{e^{2U}r'^3U'}{r}
            \nonumber \\
            &   &
            +\left(\frac{38}{3}\alpha-\frac{112}{3}\beta+\frac{16}{3}y_2\right)e^{2U}rr'U'^3
            \nonumber \\
            &   &+\left(\frac{5}{3}\alpha+\frac{8}{3}\beta-\frac{8}{3}y_2\right)e^{2U}r^2U'^4
            +(-25\alpha+60\beta-4x_2)e^{2U}r'^2U'^2  \nonumber \\
            &   &
            +\left(\frac{1}{2}\alpha+\frac{1}{4}\beta-x_3+2y_3\right)k^4\frac{e^{10U}}{r^6} +(2\beta-x_1+2y_1)\frac{e^{6U}}{r^4}k^2
             \nonumber \\
            &   & +
            \left(\frac{33}{2}\alpha-32\beta-x_2-12x_3\right)\frac{e^{6U}r'^2}{r^4}k^2
            +(-25\alpha+58\beta+24x_3+16y_3)\frac{e^{6U}r'U'}{r^3}k^2
            \nonumber \\
            &   & +\left(\frac{31}{4}\alpha-25\beta-4x_3+2y_2-24y_3\right)\frac{e^{6U}U'^2}{r^2}k^2
            \nonumber
\end{eqnarray}
This may be further simplified by reinstating the einbein $N$, and making
use of the freedom to perform field redefinitions of $N$ preserving
the property that the Lagrangian only contains powers of first
derivatives of fields:
\begin{equation}
\delta N = t_1\frac{e^{2U}}{r^2}+t_2\frac{e^{2U}r'^2}{r^2} +
t_3e^{2U}U'^2+ t_4\frac{e^{2U}r'U'}{r}+t_5\frac{e^{6U}}{r^4}k^2
\label{redN1}
\end{equation}
The Lagrangian becomes
\begin{eqnarray}
\mathcal{L} & = & 2+2r'^2-2r^2U'^2+\frac{e^{4U}}{2r^2}k^2 + k
\sigma'
\label{l1genfin} \\
            &   & +(-2\alpha+4\beta+2t_1)\frac{e^{2U}}{r^2}
            +(16\alpha-24\beta-4x_1-2t_1+2t_2)\frac{e^{2U}r'^2}{r^2} \nonumber \\
            &   & +(-32\alpha+48\beta+8x_1+8y_1+2t_4)\frac{e^{2U}r'U'}{r}
            +(16\alpha-24\beta-4x_1-8y_1+2t_1+2t_3)e^{2U}U'^2 \nonumber \\
            &   & +\left(-6\alpha+28\beta-\frac{4}{3}x_2-2t_2\right)\frac{e^{2U}r'^4}{r^2}
            +
            \left(28\alpha-\frac{152}{3}\beta+\frac{8}{3}x_2-2t_4\right)\frac{e^{2U}r'^3U'}{r}
            \nonumber \\
            &   &
            +\left(\frac{38}{3}\alpha-\frac{112}{3}\beta+\frac{16}{3}y_2+2t_4\right)e^{2U}rr'U'^3
            \nonumber  \\
            &   &+\left(\frac{5}{3}\alpha+\frac{8}{3}\beta-\frac{8}{3}y_2+2t_3\right)e^{2U}r^2U'^4
            +(-25\alpha+60\beta-4x_2+2t_2-2t_3)e^{2U}r'^2U'^2  \nonumber \\
            &   &
            +\left(\frac{1}{2}\alpha+\frac{1}{4}\beta-x_3+2y_3+\frac{1}{2}t_5\right)k^4\frac{e^{10U}}{r^6}
            +\left(2\beta-x_1+2y_1+\frac{1}{2}t_1+2t_5\right)\frac{e^{6U}}{r^4}k^2
             \nonumber \\
            &   & +
            \left(\frac{33}{2}\alpha-32\beta-x_2-12x_3+\frac{1}{2}t_2-2t_5\right)\frac{e^{6U}r'^2}{r^4}k^2
            \nonumber \\
            &   &
            +\left(-25\alpha+58\beta+24x_3+16y_3+\frac{1}{2}t_4\right)\frac{e^{6U}r'U'}{r^3}k^2
            \nonumber \\
            &   & +\left(\frac{31}{4}\alpha-25\beta-4x_3+2y_2-24y_3+\frac{1}{2}t_3+2t_5\right)\frac{e^{6U}U'^2}{r^2}k^2
            \nonumber
\end{eqnarray}
Remarkably, there exists a unique choice of the field redefinition ambiguities
such that $\mathcal{L}$ reduces to its tree-level answer,
\be
x_1 = \frac{11}{4}\alpha-5\beta \ ,\quad
x_2 = -\frac{9}{4}\alpha+7\beta \ ,\quad
x_3 = \frac{1}{16}(17\alpha-36\beta)\nn
\ee
\be
y_1 = -\frac{3}{2}\alpha+3\beta\ ,\quad
y_2 = -\frac{13}{2}\alpha+13\beta\ ,\quad
y_3 = \frac{1}{8}(-3\alpha+2\beta)\ ,\quad \label{FRsimplag}
\ee
\be
t_1 = \alpha-2\beta\ ,\quad
t_2 =-\frac{3}{2} \ ,\quad
t_3 = -\frac{19}{2}\alpha+16\beta \ ,\quad
t_4 = 11\alpha-16\beta, \quad
t_5 = \frac{21}{8}\alpha-6\beta \nn
\ee
We conclude that in the stationary, spherically symmetric sector,
$R^2$ corrections to Einstein  gravity can be completely eliminated by field
redefinitions. In particular, the Ehlers symmetry is unbroken at this order.
This result could have been anticipated \cite{Colonnello:2007qy}
from the fact that,
using field redefinitions of the four-dimensional graviton
of the form $\delta g_{\mu\nu}=\kappa_1 R_{\mu\nu}+ \kappa_2 g_{\mu\nu} R$,
the $R^2$ corrections can always be related to the Gauss-Bonnet
density, which is a total derivative. Clearly, higher order corrections
such as $R^4$ cannot not be eliminated in the same fashion. Such terms
have been discussed in \cite{Bao:2007er}.

\subsection{Einstein-Liouville Gravity}
We now consider Einstein gravity coupled to a scalar field $\phi$,
and allow an arbitrary dependence of the $R^2$ couplings in \eqref{S4}
on $\phi$,
\begin{equation} \label{SEL}
S  =  \int d^4x \sqrt{-g} \left[ R_4 +\frac{1}{2}(\partial\phi)^2
+ \alpha(\phi) \,( [R^{(4)}_{\mu\nu}]^2 - [R^{(4)}]^2 ) +
\beta(\phi) \, [R^{(4)}]^2 + \gamma(\phi)R_{GB}^2 \right]
\end{equation}
In particular, the term proportional to the Gauss-Bonnet density is no
longer a total derivative, and we no longer expect to be able to
remove all higher derivative corrections by field redefinitions.
In the rest of this section we will refrain from displaying the $\phi$
dependence, and will indicate $\phi$-derivatives with a subscript, e.g.
$\alpha_\phi \equiv d\alpha/d\phi$.

We now proceed as in section \ref{einstein}, by first performing field 
redefinitions and integrations by parts such that only powers of first 
derivatives appear in the Lagrangian. The most general field redefinition
of $r$ and $U$ compatible with these requirements is again (\ref{FRgen}),
while the field redefinition of $N$ must be generalized to 
\begin{eqnarray}
\delta N & = &
t_1\frac{e^{2U}}{r^2}+t_2\frac{e^{2U}r'^2}{r^2} +
t_3e^{2U}U'^2+ t_4\frac{e^{2U}r'U'}{r}
 \\
 &  & +t_5\frac{e^{6U}}{r^4}k^2 + t_6e^{2U}\phi'^2 +
 t_7e^{2U}U'\phi' + t_8\frac{e^{2U}r'\phi'}{r} \nonumber
\end{eqnarray}
and an extra field redefinition of $\phi$ must be introduced,
\be
\delta \phi =
(5\alpha-8\beta)\frac{e^{2U}r'\phi'}{r} +
(-4\alpha+8\beta)e^{2U}U'\phi'
 +z_1\frac{e^{2U}}{r^2}+z_2e^{2U}\phi'^2
 +z_3\frac{e^{6U}}{r^4}k^2
\ee
Moreover, all coefficients $x_i,y_i,z_i,t_i$ may now depend on $\phi$.
The full Lagrangian reads, after dropping a total derivative:
\begin{eqnarray}
\mathcal{L} & = & 2 + \frac{1}{2}r^2\phi'^2 + 2r'^2 -2r^2U'^2 + \frac{e^{4U}}{2r^2}k^2 + k \sigma'\label{monsterlag} \\
 &  &
+\left(-2\alpha+4\beta+2t_1\right)\frac{e^{2U}}{r^2}
+\left(16\alpha-24\beta+2t_1+2t_3-4x_1-8y_1\right)e^{2U}U'^2
 \nonumber  
 \eea \bea
 &  & +\left(32\alpha+48\beta+2t_4+8x_1+8y_1\right)\frac{e^{2U}r'U'}{r} \nonumber \\
 &  &
 +\left(16\alpha-24\beta-2t_1+2t_2-4x_1\right)\frac{e^{2U}r'^2}{r^2}
 +\left(-\frac{1}{2}t_1+2t_6+x_1+z_{1,\phi}\right)e^{2U}\phi'^2 \nonumber \\
 &  & +\left(-12\alpha_{\phi}+16\beta_{\phi}+2t_8+4x_{1,\phi}+\frac{4}{3}x_{2,\phi}-2z_1\right)\frac{e^{2U}r'\phi'}{r} \nonumber \\
 &  & +\left(4\alpha_{\phi}-8\beta_{\phi}+8\gamma_{\phi}+2t_7-4y_{1,\phi}+2z_1\right)e^{2U}U'\phi' \nonumber \\
 &  &
 +\left(\frac{5}{2}\alpha_{\phi}-4\beta_{\phi}-\frac{1}{2}t_8-\frac{4}{3}z_2\right)e^{2U}r'\phi'^3
 +\left(-2\alpha_{\phi}+4\beta_{\phi}-\frac{1}{2}t_7+\frac{2}{3}z_2\right)e^{2U}U'\phi'^3
 \nonumber \\
 &  & +\left(-\frac{15}{2}\alpha+12\beta-\frac{1}{2}t_2-2t_6+x_2\right)e^{2U}r'^2\phi'^2
 +\left(4\alpha_{\phi}-16\beta_{\phi}-2t_8\right)\frac{e^{2U}r'^3\phi'}{r}\nonumber
 \\
 &  & +\left(-\frac{15}{4}\alpha+7\beta-\frac{1}{2}t_3+2t_6\right)e^{2U}r^2U'^2\phi'^2 \nonumber \\
 &  & +\left(-25\alpha+60\beta+2t_2-2t_3-4x_2\right)e^{2U}r'^2U'^2 \nonumber \\
 &  &
 +\left(-\alpha_{\phi}-4\beta_{\phi}+16\gamma_{\phi}+2t_8\right)e^{2U}rr'U'^2\phi'
 +\left(8\alpha-14\beta-\frac{1}{2}t_4\right)e^{2U}rr'U'\phi'^2 \nonumber  \\
 &  & +\left(-10\alpha_{\phi}+20\beta_{\phi}-8\gamma_{\phi}-2t_7\right)e^{2U}r'^2U'\phi'
+\left(\frac{38}{3}\alpha-\frac{112}{3}\beta+2t_4+\frac{16}{3}y_2\right)e^{2U}rr'U'^3 \nonumber \\
 &  & +\left(\frac{4}{3}\alpha_{\phi}-\frac{8}{3}\beta_{\phi}-8\gamma_{\phi}+2t_7-\frac{4}{3}y_{2,\phi}\right)e^{2U}U'^3\phi' \nonumber \\
 &  & +\left(28\alpha-\frac{152}{3}\beta-2t_4+\frac{8}{3}x_2\right)\frac{e^{2U}r'^3U'}{r}
 +\left(-6\alpha+\frac{28}{3}\beta-2t_2-\frac{4}{3}x_2\right)\frac{e^{2U}r'^4}{r^2} \nonumber \\
 &  & +\left(-\frac{1}{2}t_6+\frac{1}{3}z_{2,\phi}\right)e^{2U}r^2\phi'^4
  +\left(\frac{5}{3}\alpha+\frac{8}{3}\beta+2t_3-\frac{8}{3}y_2\right)e^{2U}r^2U'^4
 \nonumber \\
 &  & +\left(2\beta+\frac{1}{2}t_1+2t_5-x_1+2y_1\right)\frac{e^{6U}}{r^4}k^2
+\left(-\frac{1}{2}t_5+\frac{1}{2}t_6+x_3+z_{3,\phi}\right)\frac{e^{6U}\phi'^2}{r^2}k^2 \nonumber \\
 &  & +\left(\frac{31}{4}\alpha-25\beta+\frac{1}{2}t_3+2t_5-4x_3+2y_2-24y_3\right)\frac{e^{6U}U'^2}{r^2}k^2 \nonumber \\
 &  & +\left(-25\alpha+58\beta+\frac{1}{2}t_4+24x_3+16y_3\right)\frac{e^{6U}r'U'}{r^3}k^2 \nonumber \\
 &  &
 +\left(\frac{33}{2}\alpha-32\beta+\frac{1}{2}t_2-2t_5-x_2-12x_3\right)\frac{e^{6U}r'^2}{r^4}k^2\nonumber
 \\
 &  & +\left(-\frac{9}{2}\alpha_{\phi}+8\beta_{\phi}-4\gamma_{\phi}+\frac{1}{2}t_8+4x_{3,\phi}-4z_3\right)\frac{e^{6U}r'\phi'}{r^3}k^2
 \nonumber  \\
 &  &
 +\left(\alpha_{\phi}-4\beta_{\phi}+10\gamma_{\phi}+\frac{1}{2}t_7-4y_{3,\phi}+6z_3\right)\frac{e^{6U}U'\phi'}{r^2}k^2
 \nonumber \\
 &  &
 +\left(\frac{1}{2}\alpha+\frac{1}{4}\beta+\frac{1}{2}t_5-x_3+2y_3\right)\frac{e^{10U}}{r^6}k^4
 \nonumber
\end{eqnarray}
While there is no longer any choice of the coefficients $x_i,y_i,z_i,t_i$ 
which removes all higher derivative corrections, enforcing the same field 
redefinitions as in (\ref{FRsimplag}) (with all coefficients being 
now functions of $\phi$) ensures that all higher derivative contributions 
become proportional to $\phi'$. This leaves six field redefinition ambiguities,
which may be used e.g. to eliminate terms of cubic and quartic order in
$U'$ and $k$:
\begin{eqnarray}
z_1 = -8\gamma_{\phi} &\quad , \quad&
z_2 = -\frac{3}{4}(\alpha_{\phi}-2\beta_{\phi}+2\gamma_{\phi}), \nonumber \\
z_3 = -2\gamma_{\phi} &\quad , \quad&
t_6 = -\frac{1}{2}\beta-z_{3,\phi}, \label{FRsimplagliouville} \\
t_7 = -5\alpha_{\phi}+10\beta_{\phi}+4\gamma_{\phi} &\quad ,
\quad& t_8 = \frac{1}{2}\alpha_{\phi}+2\beta_{\phi}-8\gamma_{\phi}
\nonumber
\end{eqnarray}
Finally, we perform the Legendre transform over $k$, obtaining
\begin{eqnarray}
L & = &
2+\frac{1}{2}r^2\phi'^2+2r'^2-2r^2U'^2-\frac{1}{2}e^{-4U}r^2\sigma'^2
\label{LagELsigma}\\
  &   &
  +\left(\frac{9}{4}\alpha-5\beta-4\gamma_{\phi\phi}\right)e^{2U}\phi'^2
+\left(\frac{1}{4}\beta-\frac{1}{4}\alpha_{\phi\phi}+\frac{1}{2}\beta_{\phi\phi}\right)e^{2U}r^2\phi'^4 \nonumber \\
  &   & +\left(\frac{13}{4}\alpha_{\phi}-7\beta_{\phi}\right)e^{2U}rr'\phi'^3
  +(-9\alpha+20\beta-4\gamma_{\phi\phi})e^{2U}r'^2\phi'^2
  +16\gamma_{\phi}\frac{e^{2U}r'^3\phi'}{r} \nonumber \\
  &   &
  +\left(\frac{5}{2}\alpha-6\beta\right)e^{2U}rr'U'\phi'^2
  -16\gamma_{\phi}e^{2U}r'^2U'\phi'
  +(\alpha-2\beta+4\gamma_{\phi\phi})e^{2U}r^2U'^2\phi'^2 \nonumber \\
  &   & +\left(-\frac{1}{4}\alpha+\frac{1}{2}\beta-\gamma_{\phi\phi}\right)e^{-2U}r^2\phi'^2\sigma'^2 \nonumber
\end{eqnarray}
In contrast to the pure gravity case, the higher-derivative terms
now break the tree-level Ehlers symmetry (\ref{sl2eh})  
explicitly\footnote{Strictly
speaking, we cannot rule out that the S-matrix computed from the
Lagrangian  \eqref{LagELsigma} preserves Ehlers symmetry, although
we find this possibility very unlikely.}.
In section \ref{ehlers}, we will show how
the symmetry under a discrete subgroup of $SL(2,\mathbb{R})$
can be restored by including the contribution of Taub-NUT instantons.

\subsection{Einstein-Maxwell Gravity}
We now study the dimensional reduction of Einstein
gravity coupled to an abelian gauge field $A$, in the presence of
higher derivative corrections of the form:
\bea
\label{smax}
S &=& \int d^4x
\sqrt{-g} \left[ R_4  -\frac{1}{4} F_{\mu\nu}^2  + \alpha
\,( [R^{(4)}_{\mu\nu}]^2 - [R^{(4)}]^2 ) + \beta \, [R^{(4)}]^2 \right.\\
&&\left.+ z_1 \,
F_{\mu\nu}F_{\nu\rho}F_{\rho\lambda}F_{\lambda\mu}
+ z_2\, (F_{\mu\nu}^2)^2+ z_3 \, \nabla^{\mu} F^{\nu\rho}
\nabla_{\mu} F_{\nu\rho}
+z_4\, R\, (F_{\mu\nu})^2  + o(\alpha')\right]\nn
\eea
where $F_{\mu\nu}=\pa_\mu A_\nu - \pa_\nu A_\mu$.
As before, we restrict to stationary spherically symmetric configurations
of the form \eqref{ds4sph}.

At two-derivative order, it is known that
the reduced Lagrangian describes the geodesic motion of a fiducial particle
on (a real cone over) the dimension 4 symmetric space $SU(2,1)/Sl(2)\times
U(1)$ \cite{Kinnersley:1977pg,Mazur:1983dc,Pioline:2006ni}.
The four scalars correspond to the scalar $U$, the time component
$\zeta$ of the Maxwell field $A_4$, the pseudo-scalar $\tzeta$
dual to the reduced Maxwell field $A_i$, and the NUT potential $\sigma $
dual to the Kaluza-Klein connection $\omega_i$. Translations along the
three axionic scalars $\zeta,\tzeta,\sigma$ are generated by three
conserved charges $q,p,k$ corresponding to the electric, magnetic and
NUT charges, respectively; they satisfy an Heisenberg algebra
$[p,q]=-2 k$, as a result of the non-trivial fibration of the
$\sigma$ direction over the $(\zeta,\tzeta)$ plane. For simplicity,
we shall restrict to {\it static} configurations with
vanishing NUT charge, $k=0$.
This allows us to express the electromagnetic field-strength directly
in terms of the conserved charges $p,q$,
\begin{equation}
F=p\,\sin{\theta}\,d\theta \wedge d\phi + i\,q \frac{e^{2U}N}{r^2}\, dt
\wedge dr
\end{equation}
The invariants are then
\begin{equation}
\sqrt{-g_4} F_{\mu\nu}^2 = 2\frac{e^{2U}N}{r^2}(p^2+q^2)
\end{equation}
\begin{equation}
\sqrt{-g_4} (F_{\mu\nu}^2)^2 = -2\frac{e^{6U}N}{r^6}(p^4+q^4)
\end{equation}
\begin{equation}
\sqrt{-g_4} F_{\mu\nu}F_{\nu\rho}F_{\rho\lambda}F_{\lambda\mu}
= 4\frac{e^{6U}N}{r^6}(p^2+q^2)^2
\end{equation}
\begin{equation}
\sqrt{-g_4} \nabla_{\mu} F_{\nu\rho}
\nabla_{\mu} F_{\nu\rho} = 12 \frac{e^{4U}(r'-rU')^2}{Nr^4}(p^2+q^2)
\end{equation}
We now proceed as in Section \ref{einstein}, performing field
redefinitions and integrations by parts so that only powers of
$U'$ and $r'$ appear in the Lagrangian: \bea \delta r & = &
x_1\frac{e^{2U}}{r} + x_2\frac{e^{2U}r'^2}{r} +
x_3\frac{e^{4U}}{r^3}(p^2+q^2) +
\left(-\frac{5}{2}\alpha+4\beta\right)e^{2U}r''\nonumber \\
 & & +(-\alpha+2\beta)e^{2U}r'U'+(2\alpha-4\beta)e^{2U}rU''
+\left(\frac{1}{4}\alpha-\beta\right)e^{2U}rU'^2 \nn \\
\delta U  &=& y_1\frac{e^{2U}}{r^2}  + y_2e^{2U}U'^2 +
+ y_3\frac{e^{4U}}{r^4}(p^2+q^2)
\left(\frac{3}{2}\alpha-3\beta\right)\frac{e^{2U}r'^2}{r^2} \\
& & + \frac{1}{2}\alpha\frac{e^{2U}r'U'}{r}-\beta e^{2U}U'' \nn
\\
\delta N & = &  t_1\frac{e^{2U}}{r^2}+t_2\frac{e^{2U}r'^2}{r^2} +
t_3e^{2U}U'^2+
t_4\frac{e^{2U}r'U'}{r}+t_5\frac{e^{4U}}{r^4}(p^2+q^2)\nonumber
\eea After dropping total derivatives the Lagrangian becomes
\begin{eqnarray}
\mathcal{L} & = & 2+2r'^2-2r^2U'^2-\frac{e^{2U}}{2r^2}(p^2+q^2)
+(-2\alpha+4\beta+2t_1)\frac{e^{2U}}{r^2}
 \label{lagEM} \\
            &   &
  +(16\alpha-24\beta-4x_1-2t_1+2t_2)\frac{e^{2U}r'^2}{r^2}
          +(-32\alpha+48\beta+8x_1+8y_1+2t_4)\frac{e^{2U}r'U'}{r}\nonumber  \\
      &  &          +(16\alpha-24\beta-4x_1-8y_1+2t_1+2t_3)e^{2U}U'^2 \nonumber  \\
            &   & +\left(-6\alpha+\frac{28}{3}\beta-\frac{4}{3}x_2-2t_2\right)\frac{e^{2U}r'^4}{r^2}
            +
            \left(28\alpha-\frac{152}{3}\beta+\frac{8}{3}x_2-2t_4\right)\frac{e^{2U}r'^3U'}{r}
            \nonumber \\
            &   &
            +\left(\frac{38}{3}\alpha-\frac{112}{3}\beta+\frac{16}{3}y_2+2t_4\right)e^{2U}rr'U'^3
            \nonumber \\
            &   & +\left(\frac{5}{3}\alpha+\frac{8}{3}\beta-\frac{8}{3}y_2+2t_3\right)e^{2U}r^2U'^4
            +(-25\alpha+60\beta-4x_2+2t_2-2t_3)e^{2U}r'^2U'^2  \nonumber  \\
            &   & +\left(-\frac{1}{2}t_1+2t_5+x_1-y_1+4z_4\right)e^{4U}{r^4}(p^2+q^2) \nonumber \\
            &   & +\left(-\frac{1}{2}t_2-2t_5+x_2-12x_3+12z_3-28z_4\right)\frac{e^{4U}r'^2}{r^4}(p^2+q^2)
            \nonumber \\
            &   & +\left(-\frac{1}{2}t_4+16x_3+16y_3-24z_3+48z_4\right)\frac{e^{4U}r'U'}{r^3}(p^2+q^2)
            \nonumber \\
            &   & +\left(-\frac{1}{2}t_3+2t_5-4x_3-y_2-16y_3+12z_3-20z_4\right)\frac{e^{4U}U'^2}{r^2}(p^2+q^2)
            \nonumber \\
            &   & +\left(-\frac{1}{2}t_5+x_3-y_3+4z_1\right)\frac{e^{6U}}{r^6}(p^2+q^2)^2
            -2z_2\frac{e^{6U}}{r^6}(p^4+q^4) \nonumber
\end{eqnarray}
Contrary to the pure gravity case, it is no longer possible to
cancel the higher-derivative corrections by appropriate choices of
the field redefinition ambiguities. A convenient choice is to set
\be
x_1 = \frac98 \alpha - \frac25 ( 7 \beta + 8 z_1 + z_3 ) \ ,\quad
x_2 =  -\frac{9}{4}\alpha+7\beta \nonumber \ ,\quad\nn
\ee
\be
x_3 = -\frac{7}{32}\alpha + \frac{11}{20}\beta - \frac45 z_1 + \frac{9}{10} z_3 -2 z_4 \nonumber
\ee
\be
y_1 =
\frac{1}{40}(5\alpha+32\beta+128z_1+16z_3) \ ,\quad
y_2 =  -\frac{13}{2}\alpha+13\beta \ ,\quad
\nn
\ee
\be
\label{emchoi}
y_3 =  -\frac{1}{20}(10\alpha-9\beta-16z_1+40z_2-12z_3+20z_4\ ,
\ee
\be
t_1 = \frac{1}{20}( 85 \alpha + 16 (-8\beta+8 z_1 + z_3)) \ ,\quad
t_2 =  -\frac{3}{2}\alpha \ ,\quad
t_3 = -\frac{19}{2}\alpha+ 16 \ ,\nn
\ee
\be
t_4 = 11 \alpha - 16 \beta \ ,\quad
t_5  =  \frac{1}{80}(45\alpha+16\beta+384z_1+48z_3-160z_4)
\nonumber
\ee
leading to the Lagrangian
\begin{eqnarray}
\mathcal{L} & = & 2+2r'^2-2r^2U'^2-\frac{e^{2U}}{2r^2}(p^2+q^2)
\\ &  &
 +\left(\frac{13}{2}\alpha-\frac{44}{5}\beta
+\frac{64}{5}z_1+\frac{8}{5}z_3\right)\frac{e^{2U}}{r^2}
+4z_2\frac{e^{6U}}{r^6}p^2q^2 \nonumber
 \nonumber \\
 &  & +(-17\alpha+24\beta-32z_2)\frac{e^{4U}r'U'}{r^3}(p^2+q^2)
 \nonumber \\
 &  & +\left(\frac{85}{4}\alpha-30\beta+32z_2\right)\frac{e^{4U}U'^2}
{r^2}(p^2+q^2) \nonumber
\end{eqnarray}
This makes it clear that higher-derivative corrections can be
eliminated only when the five couplings in the bare Lagrangian
satisfy the three relations \be \alpha=\frac{24}{17} \beta\ ,\quad
z_2=0\ ,\quad 8z_1+z_3=-\alpha/6 \ee It is of interest to study
whether the Lagrangian \eqref{lagEM} admits solutions with flat
spatial slices, as is necessary for the existence of extremal
black holes. To answer this question, one must check whether the
choices $N=1/\rho^2$ and $r=1/\rho$ are consistent with
the equations of motion of $N$ and $r$. Computation shows that
this is case only when $z_2=0$. This is in fact part of the
rationale for the choice \eqref{emchoi}, since, for general field
redefinition ambiguities, the conditions for the compatibility of
flat slices are given by
\begin{eqnarray}
z_2 &=& 0 \nonumber \\
x_1 & = &
\frac{1}{40}(-265\alpha+488\beta+20t_4-40x_2-128z_1-16z_3)
\nonumber \\
x_3 & = &
-\frac{7}{32}\alpha+\frac{11}{20}\beta-\frac{4}{5}z_1+\frac{9}{10}z_3-2z_4
 \label{condcomp} \\
t_1 & = &
\frac{33}{4}\alpha+\frac{1}{10}(-144\beta-10t_2-5t_4+64z_1+8z_3)
 \nonumber\\
t_3 & = &
-\frac{29}{4}\alpha+\frac{4}{5}(21\beta-5t_5+24z_1+3z_3-10z_4)
\nonumber \eea satisfied by \eqref{emchoi}. Thus, we find that the
assumption of the existence of extremal black holes (more
specifically, the consistency of the reduction to flat spatial
slices) requires that there should be no $(F^2)^2$ term in the
Lagrangian. It would be interesting to relate this condition to
the possibility of supersymmetrizing the Lagrangian \eqref{smax}.

\section{Ehlers Symmetry Restored\label{ehlers}}
One of the main results of the previous Section is that there exists a
choice of field redefinitions such that the one-dimensional Lagrangian
describing four-dimensional gravity in the stationary,
spherically symmetric sector involves only powers of first derivatives.
While this choice ensures the absence of spurious modes and at the same
makes the canonical quantization straightforward, it is also a particularly
convenient frame to discuss the invariance under the Ehlers symmetry
$Sl(2,\IR)$.

Indeed, returning to the Lagrangian \eqref{monsterlag} for
Einstein-Liouville gravity with general spatial slices, we may
perform the Legendre transform\footnote{Consistently with our
perturbative analysis, one should retain only the branch where
$\sigma'=- k e^{-4U}/r^2+ \mathcal{O}(\alpha')$.} over the NUT
charge $k$, so as to express the result as a function of $U'$ and
$\sigma'$. It is useful to change basis to
\begin{equation}
p = 2 i U' + e^{-2U} \sigma' \ ,\quad \bar p = -2 i U' + e^{-2U} \sigma' \ ,
\end{equation}
where $p=(d\tau/d\rho)/\tau_2$ is (the pull back of)
the left-invariant one-form on $Sl(2,\IR)/U(1)$, transforming
by a phase under the action \eqref{sl2eh} of $Sl(2,\IR)$,
\be
p\to\left(\frac{c\bar\tau+d}{c\tau+d}\right) p
\ee
This leads to the Lagrangian
\begin{eqnarray}
L & = & 2 + 2r'^2 +\frac{1}{2}r^2\phi'^2 -\frac{1}{2}r^2p\bar{p} \label{lagp} \\
  &   & +\left(\frac{9}{4}\alpha-5\beta-4\gamma_{\phi\phi}\right)e^{2U}\phi'^2
  + \frac{1}{4}\beta e^{2U}r^2\phi'^4  \nonumber \\
  &   &
  +\left(\frac{13}{4}\alpha_{\phi}-7\beta_{\phi}\right)e^{2U}rr'\phi'^3
  +\left(-9\alpha+20\beta-4\gamma_{\phi\phi}\right)e^{2U}r'^2\phi'^2
  \nonumber \\
  &   & +16\gamma_{\phi}\frac{e^{2U}r'^3\phi'}{r}
  +
  \left(-\frac{1}{4}\alpha_{\phi\phi}+\frac{1}{2}\beta_{\phi\phi}\right)e^{2U}r^2\phi'^4
  \nonumber \\
  &   &
  +i\left(-\frac{5}{8}\alpha+\frac{3}{2}\beta\right)e^{2U}rr'\phi'^2(p-\bar{p})
  +4i\gamma_{\phi} e^{2U}r'^2\phi'(p-\bar{p})
  \nonumber \\
  &   &
  +\left(-\frac{1}{8}\alpha+\frac{1}{4}\beta-\frac{1}{2}\gamma_{\phi\phi}\right)e^{2U}r^2\phi'^2(p^2+\bar{p}^2)\nonumber
\end{eqnarray}
This Lagrangian can be made invariant under a discrete
$Sl(2,\IZ)$ subgroup of the Ehlers symmetry provided any term
$a_{m,n} e^{2U} p^m \bar p^n$ is replaced by
$f_{1,m-n}(\tau,\bar\tau) p^m \bar p^n$, where $f_{1,m-n}$ is a
modular form of modular weight 0, $U(1)$ charge $m-n$ behaving as
$f_{m-n}\sim a_{m,n} e^{2U}$ in the limit where $U\to \infty$.
Such modular forms have already made an appearance in the physics
literature in discussions of the S-duality invariance of the type
IIB string in ten dimensions \cite{Green:1997me,Kehagias:1997jg}
and can be expressed as generalized non-holomorphic Eisenstein
series
\begin{equation}
f_{s,k}(\tau,\bar{\tau}) = \sum_{(p,q)\neq(0,0)}
\frac{\tau_{2}^{s}}{(p\tau+q)^{s+k}(p\bar{\tau}+q)^{s-k}}
\end{equation}
which satisfy $\bar{f}_{s,k}=f_{s,-k}$. Under modular
transformations the functions $f_{s,k}(\tau,\bar{\tau})$ transform
as:
\begin{equation}
f_{s,k}(\tau,\bar{\tau}) \to \left(
  \frac{c\tau+d}{c\bar{\tau}+d}\right)^k f_{s,k}(\tau,\bar{\tau})
\end{equation}
The leading behaviour as $U\to \infty$ uniquely selects $s=1$.
Using the identity
\be
\left( k + 2 i \tau_2 \partial_\tau \right) f_{s,k} = (s + k) f_{k+1}
\ee
and the known expression for $f_{1,0}$,
\begin{equation}
f_{1,0} = -\pi\log(\tau_2|\eta(\tau)|^4)
\end{equation}
it is easy to express $f_{s=1,k}$ for relevant values\footnote{The
Eisenstein series $f_{1,3}$ and $f_{1,4}$
would become useful if we chose not to eliminate
the cubic and quartic terms in $U'$ and $k$ in
\eqref{monsterlag}.} of $k$ in terms of the standard holomorphic
and almost holomorphic modular forms $E_4$, $E_6$ and
$\hat{E}_2=E_2-\frac{3}{\pi\tau_2}$:
\begin{equation}
f_{1,1} = \frac{\pi^2\tau_2}{3}\hat{E}_2\ ,\quad
f_{1,2} = \frac{\pi^3\tau_2^2}{18}(E_4-\hat{E}_2^2)
\end{equation}
\begin{equation}
f_{1,3} =
\frac{\pi^4\tau_2^3}{81}(2E_6-3\hat{E}_2E_4+\hat{E}_2^3)\ ,\quad
f_{1,4} =
\frac{\pi^5\tau_2^4}{324}(3E_4^2-8\hat{E}_2E_6+6\hat{E}_2^2E_4-\hat{E}_2^4)
\end{equation}
The large radius $\tau_2 \to \infty$ expansion for these functions
reads:
\begin{equation}
f_{1,k} = \frac{\pi^2}{3}\tau_2-\frac{\pi}{k} + \mathcal{O}\left(
e^{-\tau_2} \right)\ ,
\end{equation}
which also applies to $k=0$ upon replacing $1/k$ by $\log\tau_2$.
The Lagrangian \eqref{lagp} can then be covariantized
under $Sl(2,\IZ)$ into
\begin{eqnarray}
L_{cov} & = & 2 + 2r'^2 +\frac{1}{2}r^2\phi'^2 -\frac{1}{2}r^2p\bar{p} \\
  &   & +\left(\frac{9}{4}\alpha-5\beta-4\gamma_{\phi\phi}\right)\tilde{f}_{(1,0)}\phi'^2
  + \frac{1}{4}\beta\tilde{f}_{(1,0)}r^2\phi'^4  \nonumber \\
  &   &
  +\left(\frac{13}{4}\alpha_{\phi}-7\beta_{\phi}\right)\tilde{f}_{(1,0)}rr'\phi'^3
  +\left(-9\alpha+20\beta-4\gamma_{\phi\phi}\right)\tilde{f}_{(1,0)}r'^2\phi'^2
  \nonumber \\
  &   & +16\gamma'\tilde{f}_{(1,0)}\frac{r'^3\phi'}{r}
  +
  \left(-\frac{1}{4}\alpha_{\phi\phi}+\frac{1}{2}\beta_{\phi\phi}\right)\tilde{f}_{(1,0)}r^2\phi'^4
  \nonumber \\
  &   &
  +i\left(-\frac{5}{8}\alpha+\frac{3}{2}\beta\right)rr'\phi'^2\left(\tilde{f}_{(1,1)}p-\tilde{f}_{(1,1-)}\bar{p}\right)
  +4i\gamma'r'^2\phi'\left(\tilde{f}_{(1,1)}p-\tilde{f}_{(1,-1)}\bar{p}\right)
  \nonumber \\
  &   &
  +\left(-\frac{1}{8}\alpha+\frac{1}{4}\beta-\frac{1}{2}\gamma_{\phi\phi}\right)r^2\phi'^2
\left(\tilde{f}_{(1,2)}p^2+\tilde{f}_{(1,-2)}\bar{p}^2\right)\nonumber
\end{eqnarray}
with $\frac{\pi^2}{3}\tilde{f}_{s,k}=f_{s,k}$. It agrees with the
Lagrangian \eqref{lagp} from dimensional reduction in the limit
$\tau_2=e^{2U}\to \infty$, but differs by perturbative terms of
order $1/\tau_2$ and an infinite series of
exponentially suppressed terms of order $e^{-m\tau_2\pm i m \tau_1}$
The former can be
attributed to loops of gravitons running around the compact
circle, while the latter can be naturally attributed to
Taub-NUT instantons\footnote{Contributions of Taub-NUT instantons
to three-dimensional string theories have been analyzed in
\cite{Obers:2000ta}.}: the classical action of these gravitational
self-dual instantons, with topology $\IR^3 \times S^1$ at infinity,
scales as the square of the radius of the compact direction
$e^{2U}=\tau_2$ in Planck units, and includes a linear coupling
to the NUT scalar $\sigma=\tau_1$ proportional to the NUT charge.

Thus, we have given a precise realization of the proposal outlined
in \cite{Lambert:2006he}, in the toy-model of Einstein-Liouville
gravity compactified on a circle. It is clear that this procedure
works irrespective of the details of the higher-derivative action,
provided one has managed to express it in powers of first
derivatives of $U$ and $\sigma$ only (the cone variable $r$ is
spectator in this discussion). It would be interesting to study the fate of the
$SU(2,1)$ symmetry of the Einstein-Maxwell theory (or
its extension to Einstein-Maxwell-Liouville theories) along similar lines.

\section{C-map with Higher Derivative Corrections \label{cmaphd}}
In this section we give a preliminary discussion of the relation
between the higher derivative amplitudes $F_1$ and $\tilde F_1$ on
the vector and hypermultiplet branch, focussing on the simplest
case of a single universal multiplet, $n_H=1$, and no vector
multiplet $n_V=0$ in four dimensions. After dimensional reduction
to 3 dimensions, the moduli space consists of two copies of the
quaternionic-K\"ahler space $SU(2,1)/SU(2)\times U(1)$, associated
to $U,\zeta,\tzeta,\sigma$ on the vector side, and
$\varphi,\chi,\tilde\chi,a$ on the hypermultiplet side, where $S=a +
i e^{-2\varphi}$ and $Z=\chi+i \tilde\chi$ are the two chiral
multiplets in the universal hypermultiplet, $\varphi$ being the four
dimensional dilaton. For simplicity, we shall also restrict to the
$Sl(2)/U(1)$ sector of these two moduli spaces, retaining only
$(U,\sigma)$ and $(\varphi,a)$. According to
\cite{Antoniadis:1993ze,Antoniadis:1997zt}, the hypermultiplet branch in four
dimensions may receive higher-derivative corrections of the form
\be
S_{H} = \int d^4x\,\sqrt{-g_4} \left[ \frac{\nabla_\mu S\nabla^\m
\bar{S}}{S_2^2} + \tilde\alpha\, \frac{\nabla_{\mu\nu} S
\nabla^{\mu\nu} S + 
\nabla_{\mu\nu} \bar S 
\nabla^{\mu\nu} \bar S}{S_2^2} \right] \label{Shyper}
\ee
where
$\tilde\alpha$ is in general a function of all hypermultiplets,
receiving a one-loop contribution plus instanton corrections. In
line with our simplifying assumptions, and consistently with the
fact that it arises at one-loop, we shall assume that
$\tilde\alpha$ is just a constant. Moreover, we assume that the
$R^2$ couplings in four dimensions are given by \eqref{S4} where
$\alpha$ is also a constant, corresponding to a one-loop
contribution in string theory, and $\beta$ is set to zero.

Upon reducing on a stationary, spherically symmetric metric with
flat spatial slices,
\begin{equation}
ds^2 = -e^{2U}(dt+k\cos\theta
d\varphi)^2+e^{-2U}\left(\frac{d\rho^2}{\rho^4}+\frac{1}{\rho^2}d\Omega_2^2\right)
\end{equation}
and performing the now standard field redefinitions to eliminate
powers of $\varphi''$ and $a''$, the combined four-dimensional action
\eqref{S4} plus \eqref{Shyper} reduces to a one-dimensional
Lagrangian of the form
\begin{eqnarray}
\mathcal{L} & = &
-2 U'^2 - \frac12 e^{-4U}\sigma'^2
-2\varphi'^2-\frac12 e^{4\varphi}a'^2
\label{Scmap} \\
&& + \alpha\, e^{2U}\rho^4 \left( U'^4
+e^{-8U}\sigma'^4\right) + \tilde\alpha\,e^{2U}\rho^4 \left(
\varphi'^4+ e^{8\varphi}\, a'^4 \right)  \nonumber
\end{eqnarray}
where $k$ was dualized into $\sigma'$. In writing \eqref{Scmap},
we have not paid attention to the detailed form of the
higher-derivative interactions, but only exhibited their
exponential dependance on $U$ and $\varphi$.
At tree-level, $\mathcal{L}$ is invariant under
\begin{equation}
U \to - \varphi\ ,\quad \varphi \to -U \ ,\quad
\sigma \leftrightarrow a
\label{Tduality}
\end{equation}
In type II string theory, this symmetry is realized by T-duality
along the fourth circle, which exchanges the vector and
hypermultiplet branches in three dimensions. The point to be emphasized
now is that the symmetry \eqref{Tduality} is broken by the
higher-derivative
interactions in \eqref{Scmap}, unless both $\alpha$ and $\tilde\alpha$ vanish.
Indeed, a constant $\alpha$ would imply that $\tilde\alpha \sim
e^{-2\varphi-2U}$, which would correspond to a tree-level $(\nabla^2
S)^2$ contribution which vanishes in the decompactification 
limit $U\to\infty$.
Similarly, a constant $\tilde\alpha$ would imply that $\alpha \sim
e^{-2\varphi-2U}$, again a tree-level $R^2$ correction vanishing in
the decompactification limit. Either of these options would
be disastrous. Moreover, a puzzling feature of
\eqref{Scmap} is that the vector and hypermultiplet branches are
not decoupled, in contrary to common belief.

These conundrums can be simply avoided by noting that, as shown in
Section \ref{einstein},  there exists a field redefinition
scheme which removes any moduli-independent $R^2$ correction.
Similarly, there should exist field redefinitions 
on the hypermultiplet side which allow to remove any
 moduli-independent $(\nabla^2 S)^2$ coupling.
While it is not the goal of this paper to perform a systematic analysis
of this problem, we feel that the simple observation above should
serve as a word of caution
when trying to understand the relation between topological
amplitudes on the vector and hypermultiplet side.

\acknowledgments
The authors are grateful to N.~Lambert and P.~West for useful correspondence.
We also thank the authors of \cite{Colonnello:2007qy} for pointing out a
trivial claim in an earlier version of this work.
The research of B.P. is supported by the EU under contracts
MTRN--CT--2004--005104, MTRN--CT--2004--512194, and by
ANR (CNRS--USAR) contract No 05--BLAN--0079--01.

\end{document}